\documentclass[twocolumn]{revtex4-1}

\usepackage{epsfig}
\usepackage{natbib}
\usepackage{amsmath}
\usepackage{xr}

\newcommand{\myeq}[2][]{Eq#1.~(\ref{#2})}
\newcommand{\myfig}[2][]{Fig#1.~\ref{#2}}

\DeclareRobustCommand*{\myref}[2][]{%
 \begingroup
 \romannumeral-`\x % remove space at the beginning of \setcitestyle
 \setcitestyle{numbers}%
 Ref#1.~\cite{#2}%
 \endgroup
}
\newcommand{\mrm}[1]{\mathrm{#1}}
\newcommand{\unit}[1]{\,\mathrm{#1}}
\renewcommand{\Re}{R_\mrm{e}}
\newcommand{\Ve}{V_\mrm{e}}

\newcommand{\Rs}{R_\mrm{s}}
\newcommand{\dinf}{\delta_\infty}

\newcommand{\rhoL}{\rho_\mrm{L}}
\newcommand{\rhoV}{\rho_\mrm{V}}
\newcommand{\pL}{P_\mrm{L}}
\newcommand{\pV}{P_\mrm{V}}

\newcommand{\DOm}{\Delta\Omega}
\newcommand{\kB}{k_\mrm{B}}
\newcommand{\Dnc}{\Delta n^\ast}

\begin{document}

\author{Nicolas Bruot} \affiliation{Institut Lumi\`ere Mati\`ere,
  UMR5306 Universit\'e Claude Bernard Lyon 1~--- CNRS, Universit\'e de
  Lyon, Institut Universitaire de France, 69622 Villeurbanne cedex,
  France}

\author{Fr\'ed\'eric Caupin} \affiliation{Institut Lumi\`ere
  Mati\`ere, UMR5306 Universit\'e Claude Bernard Lyon 1~--- CNRS,
  Universit\'e de Lyon, Institut Universitaire de France, 69622
  Villeurbanne cedex, France}


\title{Curvature-dependence of the liquid-vapor surface tension\\
  beyond the Tolman approximation}

\begin{abstract}
  Surface tension is a macroscopic manifestation of the cohesion of
  matter, and its value $\sigma_\infty$ is readily measured for a flat
  liquid-vapor interface.  For interfaces with a small radius of
  curvature $R$, the surface tension might differ from
  $\sigma_\infty$. The Tolman equation, $\sigma (R) = \sigma_\infty /
  (1 + 2 \delta / R)$, with $\delta$ a constant length, is commonly
  used to describe nanoscale phenomena such as nucleation.
  Here we report experiments on nucleation of bubbles in ethanol and
  $n$-heptane, and their analysis in combination with their
  counterparts for the nucleation of droplets in supersaturated
  vapors, and with water data.  We show that neither a constant
  surface tension nor the Tolman equation can consistently describe
  the data. We also investigate a model including $1/R$ and $1/R^2$
  terms in $\sigma(R)$. We describe a general procedure to obtain the
  coefficients of these terms from detailed nucleation
  experiments.
  This work explains the conflicting values obtained for the Tolman
  length in previous analyses, and suggests directions for future
  work.
  \\\\
  \emph{Accepted to Physical Review Letters.}
\end{abstract}

\maketitle

Nucleation in metastable phases is an ubiquitous phenomenon, relevant
to important fields such as atmospheric research~\cite{murray12},
mechanics of plants and trees~\cite{forterre13,noblin12,wheeler08},
and in the chemical industry to avoid vapor explosions accidents (or
``spill accidents'')~\cite{debenedetti96}.  The nucleation rate is
exquisitely sensitive to the value of the surface tension $\sigma$
between the metastable and the stable phase. As the size of the
critical nucleus that triggers the phase change is in the nanometer
range, the value of $\sigma$ relevant to nucleation may differ from
the bulk one. The idea of a dependence of the surface tension on the
curvature of the interface between phases
has been studied by Tolman~\cite{tolman49}, who proposed for a
spherical droplet with radius $R$
\begin{equation}
  \frac{\sigma_\infty}{\sigma(R)}=1+\frac{2\dinf}{R}\,,
  \label{eq:Tolman}
\end{equation}
where $\dinf$ is the Tolman length.  Determining in experiments
$\sigma(R)$ is critical to developing accurate nucleation theories.
It can also serve as an input to validity checks of density functional
theory calculations and numerical simulations.  More generally, the
small scale limit at which macroscopic laws break down is an active
field of research, as shown for instance by recent studies on vapor
pressure of nanodroplets~\cite{factorovich14}, or on flows in
nanochannels~\cite{huber15}, that are of crucial importance for oil
recovery and catalysis.

The curvature dependence of surface tension has been mainly studied
theoretically and numerically, with conflicting results about the
magnitude and even the sign of the
effect~\cite{tolman49,block10,troster12,joswiak13,hruby07,bykov99,bennett12,van_giessen09,horsch12,santiso06,moody03}. The
dearth of experimental data stems from the difficulties inherent to
measurements on nanoscopic objects.  In this work we circumvent this
problem by use of the nucleation theorem
(NT)~\cite{oxtoby94,kashchiev06}, which allows obtaining information
on the nanoscopic critical nucleus from a macroscopic observable, the
nucleation rate (number of nucleation events per unit volume and
time).
In addition, we adopt a comprehensive approach, treating on the same
footing the two symmetric cases of nanodroplets (related to the
nucleation of a liquid from a supersaturated vapor, condensation) and
nanobubbles (related to the nucleation of a vapor in a metastable
liquid, cavitation).  To complement existing data on condensation, we
have performed acoustic cavitation experiments on ethanol and
heptane.

\begin{figure}
  \includegraphics{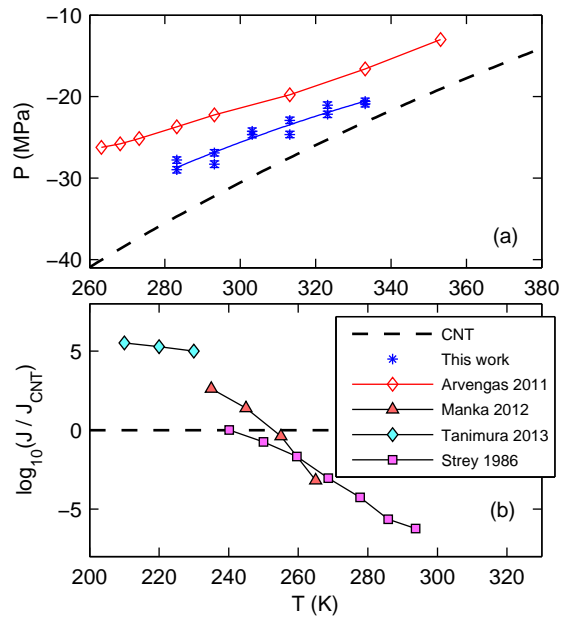}
  \caption{\label{fig:nucleation_rates} Comparison of nucleation
    pressures or rates (markers)
    with % the classical nucleation theory
    CNT$_0$ (dashed lines) in ethanol.  (a) Acoustic cavitation
    pressures obtained with the FOPH (blue stars) are compared to
    previous pressure estimates (red diamonds) via a static pressure
    method~\cite{arvengas11a}. The new, more accurate points are
    consistently more negative, as expected (see SM). The blue line is
    a guide to the eye.  (b) Condensation data.  Each graph represents
    the logarithm of the ratio between the nucleation rate and the
    CNT$_0$ prediction.  The data sources are indicated in the legend.
    See Fig.~S1 in the SM for the
    corresponding graphs for heptane and water.}
\end{figure}

In acoustic cavitation, the liquids are stretched using a few cycles
of a focused acoustic wave at 1~MHz to trigger nucleation
(see~\cite{herbert06} and Supplemental Material, SM, for details).
The wave frequency sets the experimental time and volume, and
consequently the observable nucleation rate.  The pressure at which
this rate is reached is shown in \myfig{fig:nucleation_rates}(a) for
ethanol and in Fig.~S1(a) of the SM for
heptane.  Compared to a previous study where the cavitation pressures
were based on an indirect estimate~\cite{arvengas11a}, we have now
measured them directly with a fiber-optic probe hydrophone
(FOPH)~\cite{arvengas11}.  These more accurate measurements lead to
lower pressures than the previous study, as expected because of the
nonlinearities in the acoustic wave (see SM
and~\cite{appert03,davitt10a}).

An excellent introduction to the concept of curvature-dependent
surface tension can be found in \cite{troster12}.  We just introduce here the relevant quantities on the example of
condensation.
Consider a small spherical droplet of liquid in equilibrium with
its supersaturated vapor at chemical potential $\mu$. The pressure of
the bulk liquid and vapor at $\mu$ are $\pL$ and $\pV$, and their
densities $\rho_{\text L}$ and $\rho_{\text V}$, respectively.
The key point is that the surface tension $\sigma$ depends on the
radius $R$ chosen for the dividing sphere which separates by
convention the liquid and vapor regions.
Two radii are of particular interest in
describing the droplet. The first is $\Re$, the radius of the
equimolar dividing surface. The second is $\Rs$, the radius of the
sphere at which the Laplace relation is fulfilled:
\begin{equation}
  \Delta P=\pL-\pV=\frac{2\sigma_{\text s}}{\Rs}\;,
  \label{eq:Laplace}
\end{equation}
where $\sigma_{\text s}=\sigma(R_{\text s})$. $\Rs$ allows to write
the energy barrier for nucleation $\Delta\Omega$ in a compact form:
\begin{equation}
  \label{eq:Delta_Omega}
  \DOm = \frac{4 \pi}{3} \Rs^2 \sigma_\mrm{s} =
  \frac{16 \pi \sigma_\mrm{s}^3}{3 \Delta P^2} \; .
\end{equation}

In classical nucleation theory (CNT), the nucleation rate $J$ for the
phase change is
\begin{equation}
  J=J_0\exp\left(-\frac{\DOm}{\kB T}\right)\;,
  \label{eq:J}
\end{equation}
with $J_0$ a prefactor, whose expression is given in the SM. The
knowledge of $J$ at a given $\mu$ or $\Delta P$ (the quantity
controlled in an experiment) thus gives access to $\DOm$ from
\myeq{eq:J}, $\sigma_{\text s}$ from \myeq{eq:Delta_Omega} and
$\Rs^\ast$ from \myeq{eq:Laplace}. Starred quantities are relative to
the critical nucleus, at which the energy barrier $\DOm$ is
reached. In addition, experiments can give access to $\Re^\ast$~\cite{oxtoby94,kashchiev06}.  Indeed, if the dependence of $J$ on
$\mu$ is known, the excess number of molecules in the critical nucleus
is
\begin{equation}
  \Dnc=\kB T\left(\frac{\partial\ln(J/J_0)}{\partial\mu}
  \right)_T\,.
  \label{eq:Delta_n}
\end{equation}
For $\rhoV\ll\rhoL$, and assuming a spherical critical droplet whose
density at the center reaches the bulk value, this leads to an
expression for the volume of the sphere with radius $\Re^*$ (see SM):
\begin{equation}
  \Ve^\ast=\frac{4\pi}{3}{\Re^\ast}^3=
  \frac{|\Dnc|}{\rhoL}=\frac{\kB T}{\rhoL}\left|\left(
      \frac{\partial\ln(J/J_0)}{\partial\mu}\right)_T\right|\,.
  \label{eq:V_e}
\end{equation}

\myeq[s]{eq:Delta_Omega} to \eqref{eq:V_e} hold for both cavitation
and condensation, provided that adequate expressions for $\Delta P$,
$\mu$ and $J_0$ are used.

To test different models for the surface tension, we have used
experimental values of $J$ at known $\mu$ (condensation data), or,
equivalently, $\Delta P$ at fixed $J$ (cavitation data), and $V_{\text
  e}^*$ .  The models are described below and are summarized in
Table~\ref{tab:models_summary}.

\begin{table*}
  \centering
    {\setlength\tabcolsep{6pt}%
    \begin{tabular}{c c c c}
      \toprule
      Model & Surface tension & Free parameters & Input data
      \\[8pt]
      \colrule
      CNT$_0$ & $\sigma(R_{\text s})=\sigma_\infty$ &
      None & None \\[12pt]
      CNT$_1$ & $\dfrac{\sigma_\infty}{\sigma(R_{\text s})}=1+
      \dfrac{2\delta_\infty}{R_{\text s}}$ &
      $\delta_\infty$ & $J$ \\[12pt]
      CNT$_2$ & $\dfrac{\sigma_\infty}{\sigma(R_{\text s})}=1+
      \dfrac{2\delta_\infty}{R_{\text s}}+\dfrac{\delta_\infty^2+
        \alpha}{R_{\text s}^2}$ &
      $\delta_\infty$ and $\delta_\infty^2+
      \alpha$ & $J$ and $V_{\text e}^*$ \\[8pt]
      \botrule
  \end{tabular}
  }%
  \caption{Summary of the models tested in this Letter.  For each
    model, we indicate the expression for the surface tension,
    the free parameters and the experimental data used to extract
    these parameters.}
  \label{tab:models_summary}
\end{table*}

%%%%%%%%%
% CNT_0 %
%%%%%%%%%

In the standard version of the CNT~\cite{debenedetti96a}, CNT$_0$,
$\sigma_{\text s}$ is assumed to always remain equal to the value for
a planar interface $\sigma_\infty$, which is equivalent to setting
$\Re=\Rs$, and $\DOm=16\pi{\sigma_\infty}^3/(3\Delta P^2)$.
CNT$_0$ is notorious to fail in predicting correct nucleation rates,
for cavitation~\cite{azouzi13} as well as for
condensation~\cite{ghosh10,kim04a,luijten97,holten05,manka12,bennett12,hyvarinen04,peeters02,fisk96}.
This appears clearly in \myfig[s]{fig:nucleation_rates} and
S1, where the FOPH experiments are
plotted along with condensation
data~\cite{manka12,tanimura13,strey86b,ghosh10,rudek96,holten05,wolk01,brus08},
and with a cavitation point from a water inclusion in
quartz~\cite{azouzi13}.  For instance,
\myfig[s]{fig:nucleation_rates}(b) and
S1(b,d) highlight for the three fluids a
crossover temperature below which CNT$_0$ underestimates the
condensation rates, and above which they are overestimated. This
crossover had already been observed in single data sets for ethanol
and water condensation, e.g. \cite{manka12} and~\cite{mikheev02}, and
the combination of several sets makes this conclusion stronger.

We therefore investigate other models with $\Re \neq \Rs$.
We use a functional form suggested by simulations~\cite{troster12}:
\begin{equation}
  \delta(\Rs) = \dinf + \frac{\alpha}{\Rs} \, .
  \label{eq:delta}
\end{equation}

%%%%%%%%%
% CNT_1 %
%%%%%%%%%

\begin{figure}
  \includegraphics{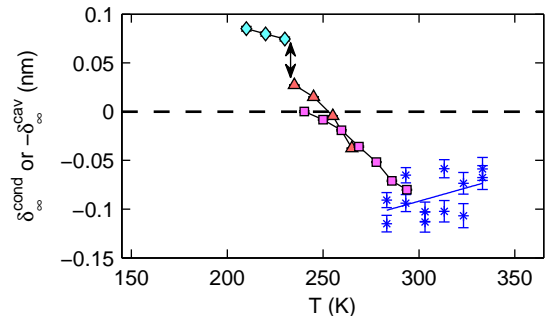}
  \caption{\label{fig:tolman_order1} $\delta_\infty$ depending on
    temperature for cavitation (star markers) and condensation (filled
    markers) in ethanol.  The legend is the same as in
    Fig.~\ref{fig:nucleation_rates}.  The double arrows point to
    discrepancies that suggest the failure of the CNT$_1$ model:
    $\delta_\infty$, which is expected to be the Tolman length, would
    not only depend on $T$.  The blue line is a guide to the eye of
    the FOPH data.  See Fig.~S2 for heptane
    and water.}
\end{figure}

Starting with a first variant of CNT, CNT$_1$, which assumes $\alpha =
0$ in \myeq{eq:delta}, it can be shown that~\cite{troster12}
\begin{equation}
  \frac{\sigma_\infty}{\sigma_\mrm{s}} = 1 + \frac{2 \dinf}{\Rs} +
  \mathcal{O} \left( \frac{1}{{\Rs}^2} \right) \, ,
  \label{eq:sigma_order1}
\end{equation}
which is similar to \myeq{eq:Tolman}. One can then calculate $\dinf$
from the experimental nucleation rates (see SM). The analysis usually
stops there~\cite{holten05,azouzi13,joswiak13}, which does not provide
a full test of CNT$_1$. We take a step further, and predict $\Ve^\ast$
from $\dinf$ with CNT$_1$ (see SM). A comparison between predicted
$V_{\text e}^*$ and $V_{\text e}^*$ deduced from the experiments with
the NT \myeq{eq:V_e} becomes possible. To our knowledge, this type of
reasoning has been employed only for water~\cite{holten05,azouzi13},
with a seemingly satisfactory agreement. By a more comprehensive
analysis of all the data sets gathered for ethanol, heptane and water,
covering a broader range for temperature and degree of metastability,
we find discrepancies that reveal the actual failure of CNT$_1$.
\myfig[s]{fig:tolman_order1} and S2 show
$\dinf$ calculated from the cavitation and condensation experiments.
Quantities relative to cavitation and condensation are labelled by
`cav' or `cond', respectively.  If CNT$_1$ were valid, we would expect
to find that $\delta_\infty^\text{cav}$ and
$\delta_\infty^\text{cond}$ do not depend on $R_{\text s}^*$
and that $\dinf^{\text{cav}} = -\dinf^{\text{cond}}$ for the same
temperatures~\cite{troster12}. For each of the fluids, the points do
not collapse on a single curve, as indicated by the double arrows,
even when taking into account the experimental uncertainties (see SM
for details). This suggests that $\delta$ does in fact depend on
$R_s^*$.
We emphasize that the disagreement can usually not be seen when
looking at the data of a single condensation experiment.
This is because various independent supersaturation and temperature
values are needed to conclude and the combination of several
condensation and cavitation experiments extends the range of both
parameters.
The crossover mentioned earlier translates here into a change of
$\delta_\text{cond}$ from positive to negative values when $T$
increases (\myfig[s]{fig:tolman_order1} and
S2). This behavior is not in support of
CNT$_1$.
The situation gets even worse when comparing in
\myfig[s]{fig:critical_volumes} and S3
$V_{\text e}^*$ extracted from this model (small, light markers) with
$V_{\text e}^*$ obtained using the NT (big, dark markers).
In overall, these agree at low temperature, but strongly disagree
above 250-300~K depending on the fluid.
We note that the approximations needed to deduce $\Ve^\ast$ from
experiments with \myeq{eq:V_e} lead to an underestimate of $\Ve^\ast$
for both cavitation and condensation, so that the disagreement with
CNT$_1$ can be only stronger than shown in
\myfig[s]{fig:critical_volumes} and S3.
Also, the critical volumes for cavitation may display a systematic
error bigger than the statistical error bars shown here because of an
extrapolation in the data analysis.  All these details are
investigated in the SM and Refs.~\cite{oxtoby98,kashchiev03} and do
not change any of our conclusions.

\begin{figure}
  \includegraphics{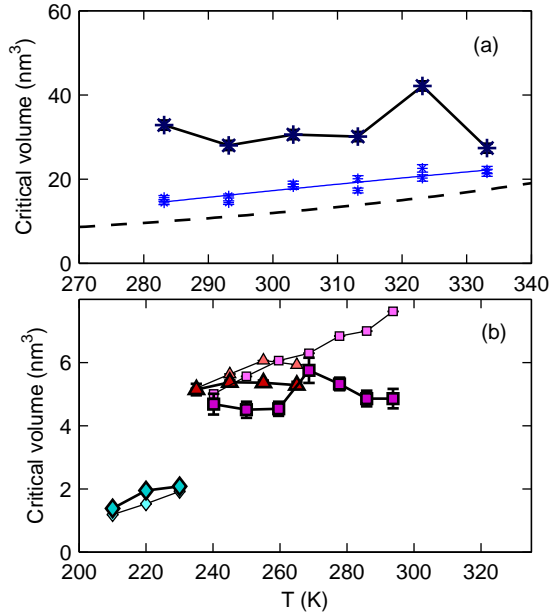}
  \caption{\label{fig:critical_volumes} Critical volumes in (a)
    cavitation and (b) condensation experiments for ethanol. The thick
    and thin markers represent the volumes from the NT and from
    CNT$_1$, respectively. The legend is the same as in
    Fig.~\ref{fig:nucleation_rates}.  See
    Fig.~S3 for heptane and water.}
\end{figure}

%%%%%%%%%
% CNT_2 %
%%%%%%%%%

\begin{figure}
  \includegraphics{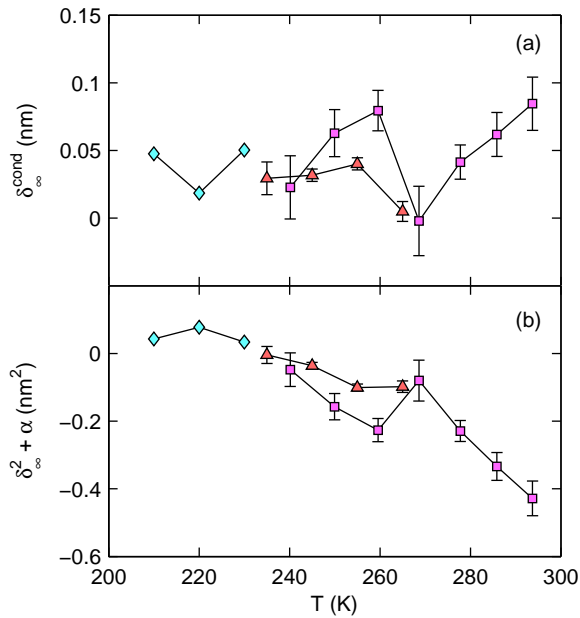}
  \caption{\label{fig:tolman_order2} CNT$_2$ parameters
    $\delta_\infty$ and $\delta_\infty^2 + \alpha$ derived from
    condensation experiments.  See \myfig{fig:nucleation_rates} for
    the legend, and Fig.~S4 for heptane and
    water.}
\end{figure}

We now move on to a second variant of CNT, CNT$_2$, based on
\myeq{eq:delta} with two parameters $\dinf$ and $\alpha \neq 0$. It
can then be shown that~\cite{baidakov14}:
\begin{equation}
  \frac{\sigma_\infty}{\sigma_\mrm{s}} = 1 + \frac{2 \dinf}{\Rs} +
  \frac{{\dinf}^2 + \alpha}{{\Rs}^2} + \mathcal{O} \left(
    \frac{1}{{\Rs}^3} \right) \, .
  \label{eq:sigma_order2}
\end{equation}
It may seem that adding an extra parameter would automatically allow a
better fitting of the data. But we also use more experimental input
(see Table~\ref{tab:models_summary}). For CNT$_1$, we used only the
experimental nucleation rates to calculate $\dinf$, and compared the
CNT$_1$ prediction for $\Ve^\ast$ with the values from experiments on
which the nucleation theorem can be applied. For CNT$_2$, we use both
the nucleation rates and $\Ve^\ast$ from those experiments to directly
calculate $\dinf$ and $\dinf^2 + \alpha$ (see SM). The success of the
approach must therefore be assessed by checking if the various data
sets lead to master curves for both $\delta_\infty$ \emph{and}
$\delta_\infty^2 + \alpha$ as a function of the temperature. The
results are plotted in \myfig[s]{fig:tolman_order2} and
S4. We have omitted here the FOPH data,
since, unfortunately, the corresponding error bars on $\delta_\infty$
and $\delta_\infty^2 + \alpha$ would be of the order of the size of
the $y$-axes (see SM).
For heptane, only one data set is available and does not allow to test
the data collapse.  For the other fluids, we observe a more
consistent description of the data (see in particular the improvement
of the discrepancy between the Tanimura 2013 and the Manka 2012
data
compared to
\myfig{fig:tolman_order1}).
The agreement is still not perfect, even when taking into account
the statistical uncertainties on the points.  This is possibly
indicative of systematic errors specific to the different experiments,
or of some limitation of the theory, such as that of a spherical
critical nucleus.
Compared to CNT$_1$, CNT$_2$ yields a Tolman length $\dinf$ with a
weaker temperature dependence, and for ethanol and water $\dinf$ now
keeps a positive sign. For heptane, $\dinf$ is close to zero, maybe
slightly negative. For the three fluids, we noted that the second
order term $(\dinf^2 + \alpha)/(R_{\text s}^*)^2$ is often of the same
order of magnitude as the first order term $\delta_\infty / R_{\text
  s}^*$ (see SM).

%%%%%%%%%%%%%%%%%%%%%%%%%%%%%
% Discussion and conclusion %
%%%%%%%%%%%%%%%%%%%%%%%%%%%%%

Our results show that, at least for ethanol, heptane and water, the
usual Tolman equation \myeq{eq:Tolman} is not enough to properly
describe experiments. Therefore, attempts to analyse experiments with
CNT$_1$ (such as in \cite{azouzi13,holten05,joswiak13}) may
yield inaccurate determinations of $\dinf$, and this study partly
explains the confusion in the longstanding debate on the sign of the
Tolman length~\cite{malijevsky12}.  The CNT$_2$ approach seems to give
more consistent results.

A variety of simulations have been realized and motivated our work.
We have tried to compare the experiments to these.
For heptane, the $\delta_\infty^\text{cond}$ we found is close to zero
and possibly negative (around $-0.02\unit{nm}$). An expression of the
Tolman length as a function of the isothermal compressibility
$\kappa_{\text T}$ has been proposed~\cite{bartell11,blokhuis06}:
$\delta_\infty = -\kappa_{\text T} \sigma_\infty$. At 265~K, the
formula yields $\delta_\infty \approx -0.03$~nm for heptane, which is
close to the experiments in
Fig.~S4(a). Heptane may be crudely
approximated by a Lennard-Jones fluid, for which DFT calculations and
MC simulations~\cite{block10,troster12} seem to point to a slightly
negative value for $\delta_\infty^\text{cond}$ ($-0.07$~nm).  However, MD simulations~\cite{baidakov14} find
a positive temperature-dependent $-\delta_\infty^\text{cav}$
($+0.1$~nm at 265~K). Also, Iwamastu~\cite{iwamatsu94} estimated the
Tolman length from the correlation lengths of the liquid and vapor
phases, which for heptane translates into $\delta_\infty^{\text{cond}}
\approx -0.2$~nm, which has a larger magnitude than the experimental
value.
The parameter $\delta_\infty^2 + \alpha$ has been estimated by MD
simulations~\cite{baidakov14} and by DFT~\cite{block10,blokhuis13}.
Rescaled to heptane, these estimates all lead to a positive
$(\delta_\infty^2 + \alpha) / R_{\text s}^2$ of about
0.4~\cite{baidakov14}, while the Rudek data set displays mostly
negative values: $(\delta_\infty^2 + \alpha) / R_{\text e}^2 = -0.1$
in average in the 250-275~K range.
By identifying \myeq{eq:sigma_order2} to the Helfrich form of the
surface free energy in \myref{helfrich73}, we find the average
curvature-elastic moduli $2 k_\text{c} + \bar k_\text{c} = 7 \times
10^{-22}$~J.

For the other fluids, the different experiments partially collapse on
master curves, thus supporting the CNT$_2$ model, but they can hardly
be compared to simulations or DFT estimates based on the Lennard-Jones
potential.
For water, we first note that within CNT$_2$, the
cavitation~\cite{azouzi13} and the condensation~\cite{brus08}
experiments yield to positive $\delta_\infty^{\text{cond}}$ or
$-\delta_\infty^{\text{cav}}$.  This sign is consistent with
simulations based on a monoatomic model of water (mW) where a
departure from the Kelvin equation is observed at small droplet
radii~\cite{factorovich14}.  However, the CNT$_2$ analysis of two
simulations with TIP4P/2005~\cite{joswiak13,gonzalez14} would give the
opposite sign: $\delta_\infty^{\text{cond}} = -0.066$~nm at 300~K
for~\cite{joswiak13} that measured directly the radius-dependence of
the surface free energy of droplets, and $-\delta_\infty^{\text{cav}}
= -0.067$~nm for the cavitation simulations in~\cite{gonzalez14} from
which we have calculated $\delta_\infty$ with the energy barrier and
the critical volume (using the data from their ``M-method'' to
estimate $V_{\text e}^*$).

While our conclusions on the inaccuracy of the CNT$_1$ model are
unambiguous ---~we strongly recommend not to use the Tolman equation
when analysing nucleation data~---, they call for further experiments
to confirm the CNT$_2$ model. As the vapor supersaturation can be
varied over a broad range in experiments on condensation, they should
be more appropriate than cavitation. For ethanol, the success of
CNT$_2$ is already very promising, and we provide in the SM overall
fitting parameters for $\delta_\infty$ and $\delta_\infty^2 + \alpha$
that can be used to predict the nucleation rate from any
condition. For other fluids, our study provides a procedure with
which future measurements of nucleation rates and critical volumes
can be analysed.

\begin{acknowledgments}
  We acknowledge funding by the ERC under the European FP7 Grant
  Agreement 240113, and by the Agence Nationale de la Recherche Grant
  09-BLAN-0404-01.
\end{acknowledgments}

\end{document}

\vspace*{2.5cm}
\title{Supplemental Material for\\
Curvature-dependence of the liquid-vapor surface tension\\
beyond the Tolman approximation}

\author{Nicolas Bruot and Fr\'ed\'eric Caupin}

\affiliation{Institut Lumi\`ere Mati\`ere, UMR5306 Universit\'e Claude Bernard Lyon 1-CNRS, Universit\'e de Lyon and Institut Universitaire de France, 69622 Villeurbanne cedex, France}

\begin{abstract}
  This note provides additional information on the experiments and the
  analysis presented in the Letter, and presents an analysis of
  $n$-heptane and water nucleation data to complement the ethanol
  data.  It also discusses the uncertainties and approximations to
  support our conclusions.  When not specified, the notations and
  legends in this document are the same as in the main text.
\end{abstract}

\maketitle

\tableofcontents

\newpage

\section{Materials and methods}

We have performed acoustic cavitation in $n$-heptane (Sigma Aldrich,
puriss. p.a., $\geq 99$\%) and ethanol (VWR Prolabo Chemicals,
$99.98$\% v/v), using a hemispherical piezoelectric transducer to
focus 1~MHz sound bursts (a few cycles long) in a small region of the
liquid far from any wall~\cite{herbert06}. Ramping the excitation
voltage of the transducer, the cavitation probability increases from 0
to 1. The ``cavitation threshold'' corresponds to a 50\% cavitation
probability during a burst.  In a previous study~\cite{arvengas11a},
the pressure at the focus was estimated indirectly by studying the
effect of the static pressure in the liquid on the cavitation
threshold.  Here, we have measured the density of the fluid at the
focus directly with a fiber-optic probe hydrophone~\cite{arvengas11},
which is sensitive to the modulation of the refractive index by the
sound wave.  To convert the density into a pressure, we used an
equation of state for the liquid at positive pressure, and
extrapolated it down to about $-30$~MPa.  More details will be given
elsewhere.

\section{Comparison between the static pressure and FOPH methods}

Figs.~1(a) and S1(a) show pressures at the cavitation
threshold slightly more negative with the FOPH than the previously
reported values. Here we give an explanation for the discrepancy.

The static pressure method in \myref{arvengas11a} was based on the
dependence of the transducer voltage at the cavitation threshold on
the positive static pressure applied to the liquid. A linear
extrapolation gave an indirect estimate of the negative cavitation
pressure. However, nonlinearities lead to extrapolated
pressures less negative than the real ones~\cite{appert03}.  The new experiments with a FOPH give direct access to the
density of the liquid at the cavitation threshold.  The only remaining
assumption resides here in the conversion of the density into a
pressure, that requires to extrapolate to negative pressure an
equation of state measured at positive pressure.  In the case of
water, we have previously measured the equation of state at negative
pressure and proven that this assumption is valid~\cite{davitt10a}.
It is reasonable to assume that the extrapolation would also be valid
for heptane and ethanol, thus yielding to FOPH points that are more
accurate than the points from the static pressure method.

\section{Study of $n$-heptane and water}

\begin{figure}[tttt]
  \centering
  \includegraphics{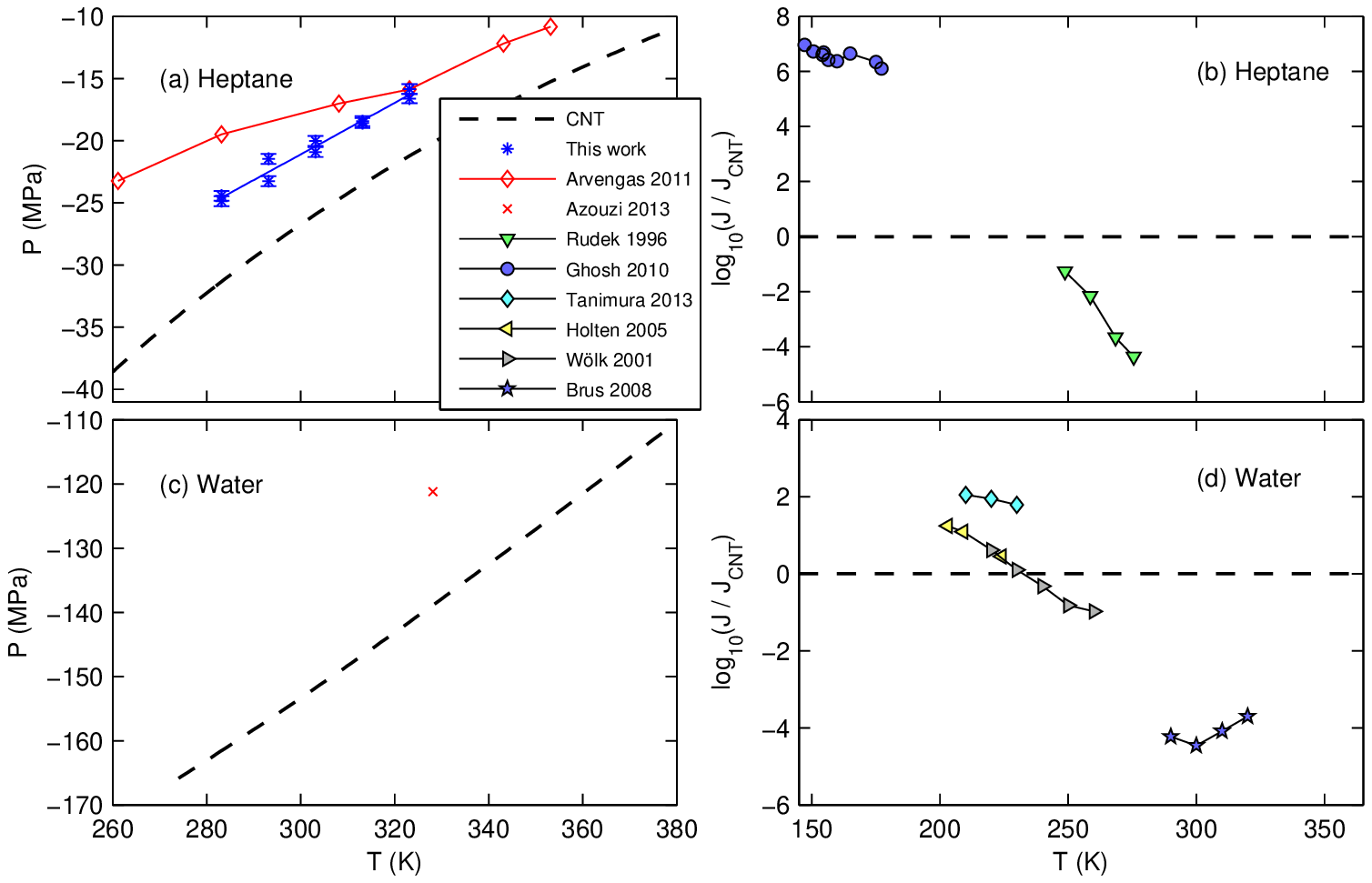}
  \caption{\label{suppl:fig:nucleation_rates_other} Same as
    Fig.~1 in the Letter, but for (a,b) heptane
    and (c,d) water.  The point in (c) corresponds to the cavitation
    pressure measurement in a water inclusion in a quartz crystal
    \cite{azouzi13}.}
\end{figure}

The analysis carried in the Letter on ethanol can be extended to other
liquids.  Figures~\ref{suppl:fig:nucleation_rates_other},
\ref{suppl:fig:tolman_order1_other}, \ref{suppl:fig:critical_volumes_other},
\ref{suppl:fig:tolman_order2_other} below show results for $n$-heptane and water.

\section{Formula of the modified CNTs}

This sections gives the formula we have used to calculate the
CNT$_1$ and CNT$_2$ parameters, and the critical volumes from the
nucleation theorem.

\subsection{CNT$\boldsymbol{_1}$}

In the CNT$_1$, Eq.~(8) is written at the critical
radius, $R_{\text s} = R_{\text s}^*$.  $R_{\text s}^*$ is trivially
deduced from the Laplace equation as a function of the energy barrier
$\Delta \Omega$, and $\Delta P$, that are known from the experiments:
\begin{equation}
  \label{suppl:eq:R_s}
  R_{\text s}^* = \left( \frac{3 \Delta \Omega}{2 \pi \Delta p}
  \right)^{1/3}
\end{equation}
Since Eq.~(3) with $R_{\text s} = R_{\text s}^*$ links
$\sigma_{\text s}(R_{\text s}^*)$ to $\Delta \Omega$ and $R_{\text
  s}^*$, we obtain the following expression for the Tolman length:
\begin{equation}
  \delta_{\infty} = \frac{\sigma_\infty}{\Delta P} \left[ 1 - \left(
    \frac{3 \Delta \Omega}{16 \pi \sigma_\infty^3} \right)^{1/3} \right]
\end{equation}

Compared to CNT$_0$, the critical equimolar radius now depends on
$\delta_\infty$.  The derivation is done in \myref{holten05} and gives
\begin{equation}
  \label{suppl:eq:V_e_order1}
  R_{\text e}^* = \frac{2 \sigma_\infty}{\Delta P} - \delta_\infty
\end{equation}

\afterpage{\clearpage}

\begin{figure}[bbbb]
  \centering
  \includegraphics{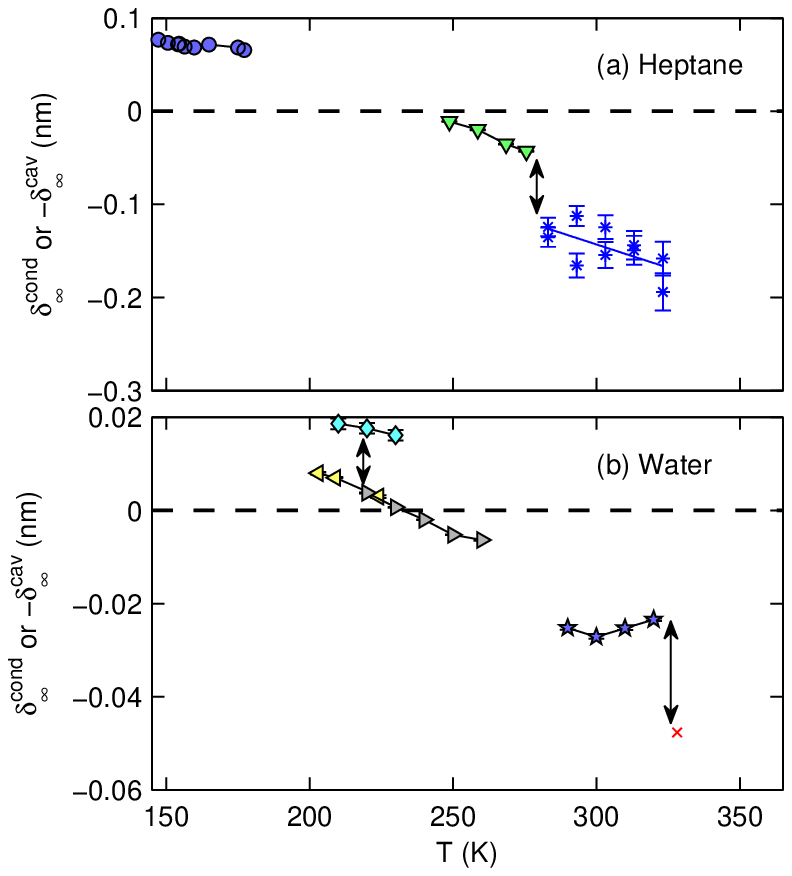}
  \caption{\label{suppl:fig:tolman_order1_other} Same as
    Fig.~2, but for (a) heptane and (b) water.}
\end{figure}

\pagebreak

\subsection{CNT$\boldsymbol{_2}$}

Eqs.~(2), (7) and (9)
form a system of equations that can be solved for $\delta_\infty$ and
$\delta_\infty^2 + \alpha$.  The relevant solutions are
\begin{eqnarray}
  \label{suppl:eq:order2_sols}
  \left\{
  \begin{array}{rcl}
    \delta_\infty &=& - \dfrac{R_{\text s}^*}{2}
    (1 - \sqrt \Delta) \\
    \delta_\infty^2 + \alpha &=& \left( R_{\text s}^* \right)^2
    \left( \dfrac{2 \sigma_\infty}{R_{\text s}^* \Delta P} - \sqrt
    \Delta \right)
  \end{array}
  \right.
\end{eqnarray}
where
\begin{equation}
 \Delta = 1 - \frac{4 R_{\text e}^*}{R_{\text s}^*} + \frac{8
 \sigma_\infty}{R_{\text s}^* \Delta P}
\end{equation}
Here, $R_{\text e}^*$ is obtained in the experiments from the
nucleation theorem and $R_{\text s}^*$ is given by \myeq{suppl:eq:R_s}.

\afterpage{
\begin{figure}
  \centering
  \includegraphics{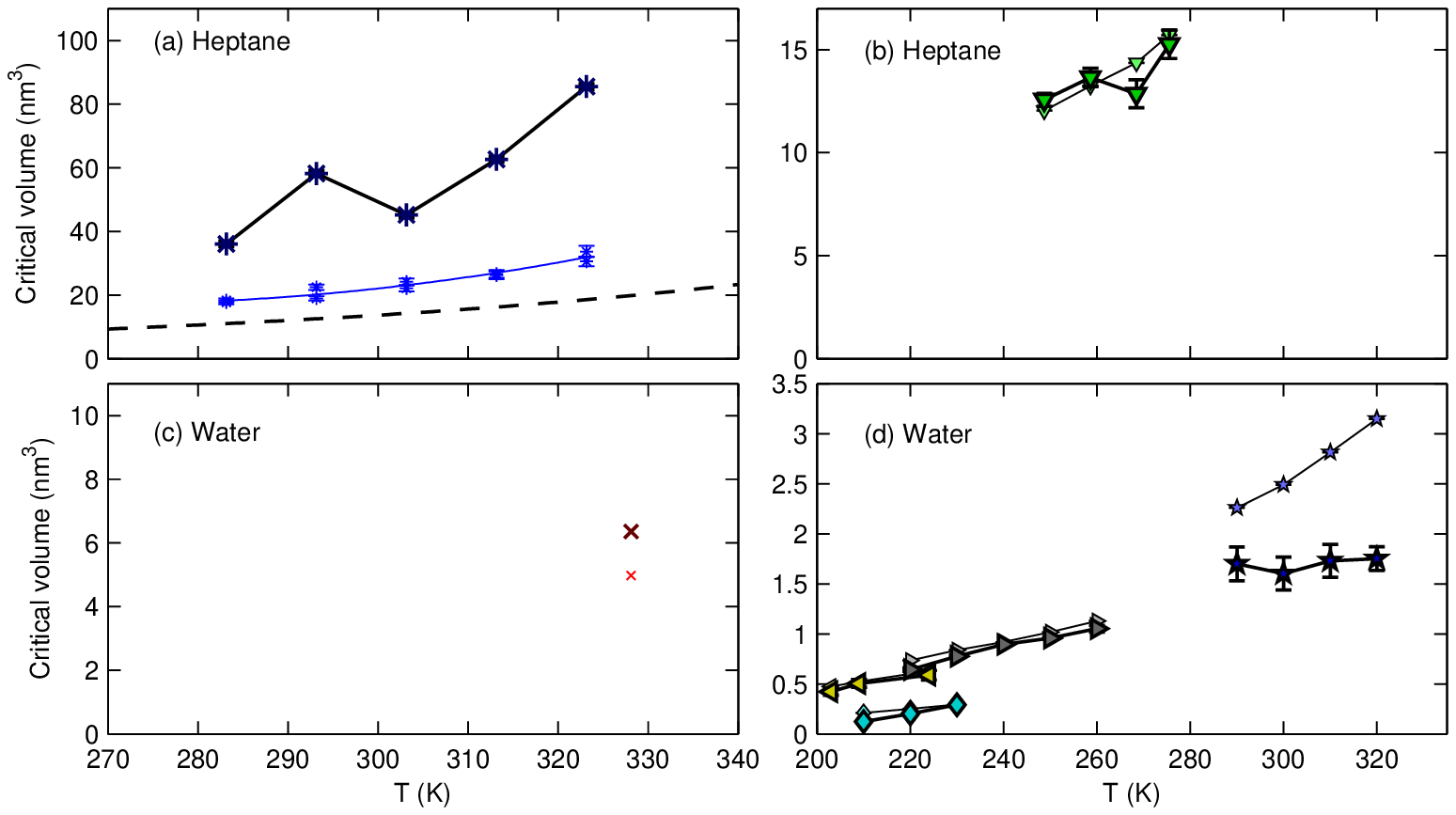}
  \caption{\label{suppl:fig:critical_volumes_other} Same as
    Fig.~3, but for (a,b) heptane and (c,d)
    water.}
\end{figure}

\begin{figure}
  \centering
  \includegraphics{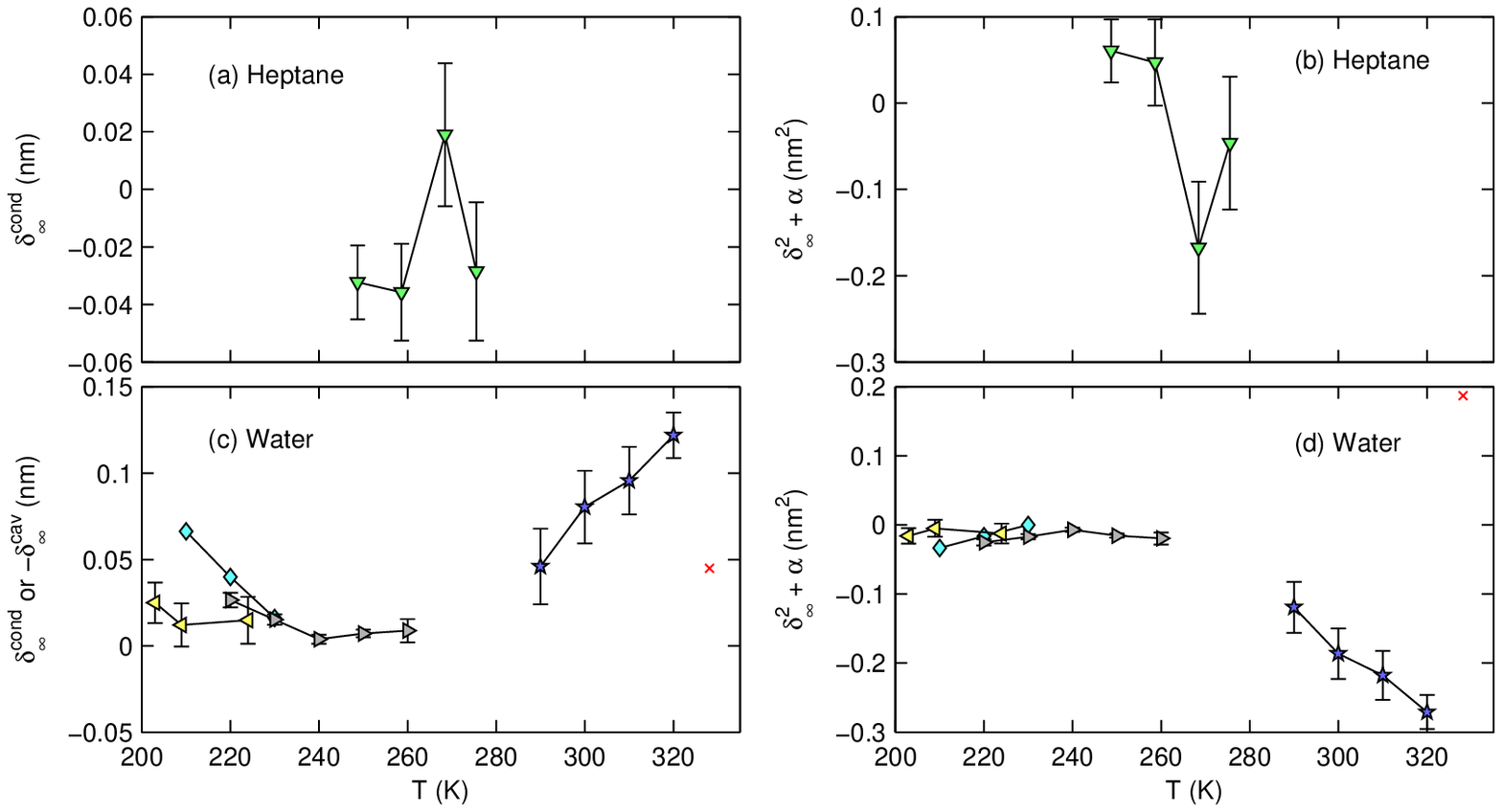}
  \caption{\label{suppl:fig:tolman_order2_other} Same as
    Fig.~4, but for (a,b) heptane and (c,d) water.}
\end{figure}
}

\subsection{Choice for the kinetic prefactor}
\label{suppl:sec:nucl_thres}

\subsubsection{Cavitation}

The CNT$_0$ and its variants CNT$_1$ and CNT$_2$ rely on the choice of
an expression for the kinetic prefactor $J_0$.

For cavitation, we chose $J_0 V \tau = 10^{19}$, where $V$ is the
volume where the acoustic wave is focalized, and $\tau$ the duration
of an acoustic burst.  The value is taken from our previous study in
water~\cite{herbert06}.  The actual value of $J_0 V \tau$ in the
present experiments might differ from the $10^{19}$ value.  However, a
change by a factor 10 in this constant only leads to a shift of the
experimental points
by 
about 0.015~nm for the CNT$_1$'s $\delta_\infty$ and 0.5~nm$^3$ for
$V_\text{e}^*$ (for both heptane and ethanol).  This is much smaller
than the statistical deviations seen by repeating cavitation pressure
measurements at the same temperature several times.

\subsubsection{Condensation}

For condensation, the nucleation rates are calculated from the
supersaturation $S = P_{\text v} / P_{\text{sat}}(T)$, where $P_{\text
  v}$ is the pressure of the metastable vapor, and $P_{\text{sat}}(T)$
is the equilibrium vapor pressure for a flat interface.  Treating the
vapor as a perfect gas, and the liquid as an incompressible phase
leads to:
\begin{equation}
  \Delta P = \frac{k T \ln S}{v_{\text l}}
\end{equation}
where $k$ is the Boltzmann constant and $v_{\text l}$ the volume per
molecule in the liquid.  (Including gas non-idealities has been shown
to have little effect on the nucleation rates for
$n$-nonane~\cite{fisk96}.)
For the kinetic prefactor, we used
\begin{equation}
  \label{suppl:eq:J_0_cond}
  J_0 = \sqrt{\frac{2 \sigma_\infty}{\pi m_{\text l}}} v_{\text l} \,
  S^2 \left( \frac{P_{\text{sat}}(T)}{k T} \right)^2
\end{equation}
with $m_{\text l}$ the mass of a molecule.  The actual value of $J_0$
is still being debated.  In particular, a ``$1/S$ correction'' is
sometimes added to \myeq{suppl:eq:J_0_cond}~\cite{holten05}.  This, again,
only leads to insignificant changes in the quantities explored in this
study.  For instance, the typical shifts from the data
in~\cite{tanimura13} are: 2 orders of magnitude for $J_{\text{exp}} /
J_{\text{CNT}}$, 0.1~nm$^3$ for $V_\text{e}^*$, 0.02~nm for the
CNT$_1$'s $\delta_\infty$, and $2 \times 10^{-4}$~nm and
$0.015$~nm$^2$ for the CNT$_2$'s $\delta_\infty$ and $\delta_\infty^2
+ \alpha$.

\subsection{Derivation of the critical volumes from the nucleation
  theorem}

Critical volumes are obtained from the nucleation theorem
Eq.~(5).  When writing Eq.~(6) and converting $\Delta
n^*$ into a volume, two consecutive approximations are made.  First,
we assume that the density at the center of the nucleus is the density
of the homogeneous phase, $\rho_{\text L}$ or $\rho_{\text V}$.  This
allows us to link the critical volume to the excess number of
molecules by:
\begin{eqnarray}
  \label{suppl:eq:V_e_n_e_droplets}
  V_{\text e}^* &=& \frac{n_{\text e}^*}{\rho_{\text L}} =
  \frac{\Delta n^*}{\rho_{\text L} - \rho_{\text V}}
  \quad \text{for droplets} \\
  \label{suppl:eq:V_e_n_e_bubbles}
  &=& \frac{n_{\text e}^*}{\rho_{\text V}} = - \frac{\rho_{\text
      V}}{\rho_{\text L} - \rho_{\text V}} \Delta n^* \quad
  \text{for bubbles}
\end{eqnarray}
where $n_{\text e}^*$ is the number of molecules in the nucleus.  The
second approximation we make is that $\rho_{\text L} \gg \rho_{\text
  V}$, which is easily satisfied: for the data analyzed in the Letter
and here, the maximal value of $\rho_{\text V} / \rho_{\text L}$ is
$0.68 \times 10^{-3}$ for condensation and $1.1 \times 10^{-3}$ for
the FOPH experiments.  The critical volume then simply becomes
\begin{equation}
  V_{\text e}^* = \frac{|\Delta n^*|}{\rho_{\text L}}
\end{equation}
for droplets and bubbles.

The two approximations above lead to an underestimate of the critical
volumes~\cite{oxtoby98,kashchiev03}.  Correcting for these would
therefore make the discrepancy between the real critical volumes and
the CNT$_1$'s volumes that we highlight in the main text stronger.

\section{Effect of the extrapolation of the voltage to pressure
  relation (cavitation experiments)}
\label{suppl:sec:p_u_extrapol}

\begin{figure}
  \centering
  \includegraphics{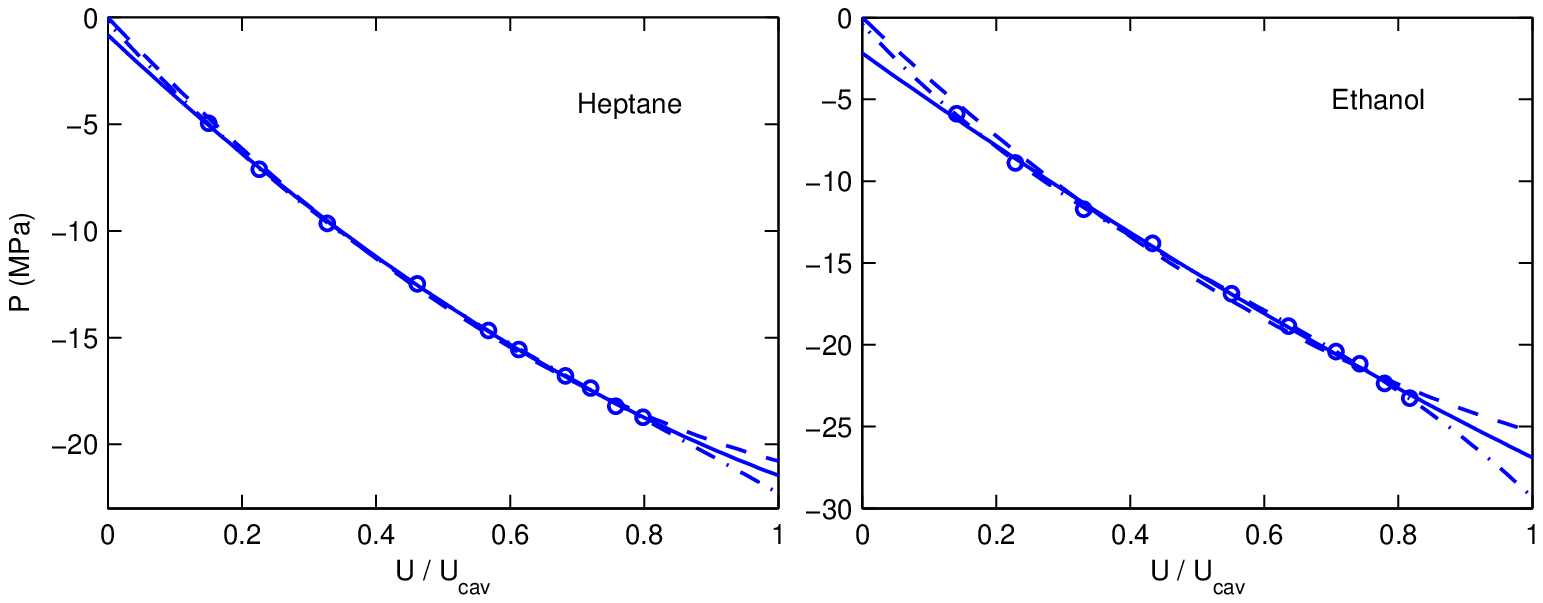}
  \caption{\label{suppl:fig:P_U_fits_all_extrapols} Pressure depending on
    the amplitude of the voltage on the piezo-electric transducer for
    $T = 293$~K.  The markers represent the measurements and the lines
    correspond to the different polynomials used to extrapolate the
    curve up to $U = U_\text{cav}$: $P_1$ (dashed line), $P_2$ (solid
    line) and $P_3$ (dash-dotted line).}
\end{figure}

A large part of the uncertainties in our experiments do not come from
statistical error bars.  To measure the cavitation pressures and the
critical volumes in the fiber-optic probe hydrophone experiments,
pressures are measured for different amplitudes of the sound wave
created by a piezo-electric transducer.  The amplitude is controlled
by the amplitude of the oscillatory voltage $U$ applied to the
transducer (see \myref{arvengas11} for details on the setup).
Unfortunately, the voltage cannot be increased up to the value for
which there would be 50~\% chance to cavitate as it would damage the
end facet of the fiber.  To obtain the pressure $P_\text{cav}$ at the
voltage $U_\text{cav}$, the pressure is measured for several values of
$U$ below about $0.8 U_\text{cav}$, and we fit the data with some
function to extrapolate the pressures to $P_\text{cav}$.  We have
tried several functions for the extrapolation, and three of them gave
sufficiently small residuals:
\begin{equation}
  P_1(U) = a_2 U^2 + a_1 U \, ,
\end{equation}
\begin{equation}
  P_2(U) = b_2 U^2 + b_1 U + b_0
\end{equation}
and
\begin{equation}
  P_3(U) = c_3 U^3 + c_2 U^2 + c_1 U + c_0
\end{equation}
where the $a_i$, $b_i$ and $c_i$ are the fitting coefficients.  We
show the three extrapolations for a given temperature in heptane and
ethanol in \myfig{suppl:fig:P_U_fits_all_extrapols}.  On this plot,
$P_\text{cav}$ can simply be read for $U = U_\text{cav}$.  To
calculate the critical volumes from the nucleation theorem within the
framework of CNT$_1$, Eq.~(6) is rewritten in terms of a
derivative $\partial P / \partial U$ of the pressure with the voltage,
so that the critical volumes are related to the slope of $P(U)$ at
$U_\text{cav}$.
Since we could not find any strong argument to determine which
extrapolation is the best, we display in
\myfig[s]{suppl:fig:cavitation_pressures_all_extrapols},
\ref{suppl:fig:tolman_order1_all_extrapols} and
\ref{suppl:fig:critical_volumes_all_extrapols} the quantities obtained with
the three polynomials.  The graphs of the main text (and their
equivalents for heptane and water in
\myfig[s]{suppl:fig:nucleation_rates_other}, \ref{suppl:fig:tolman_order1_other},
\ref{suppl:fig:critical_volumes_other} and \ref{suppl:fig:tolman_order2_other})
use polynomial $P_2$ as its residuals were slightly better than for
the other functions and because it often lies between the values
computed with $P_1$ and $P_3$.  The choice of the function for the
extrapolation can lead to significant changes of the various
quantities plotted, especially the critical volumes.  However, no
matter what function is used, there is no master curve emerging in
\myfig{suppl:fig:tolman_order1_all_extrapols} (for heptane), and it appears
very unlikely from \myfig{suppl:fig:critical_volumes_all_extrapols} that the
real critical volumes from the nucleation theorem could match the
volumes from the CNT$_1$.

\begin{figure}
  \centering
  \includegraphics{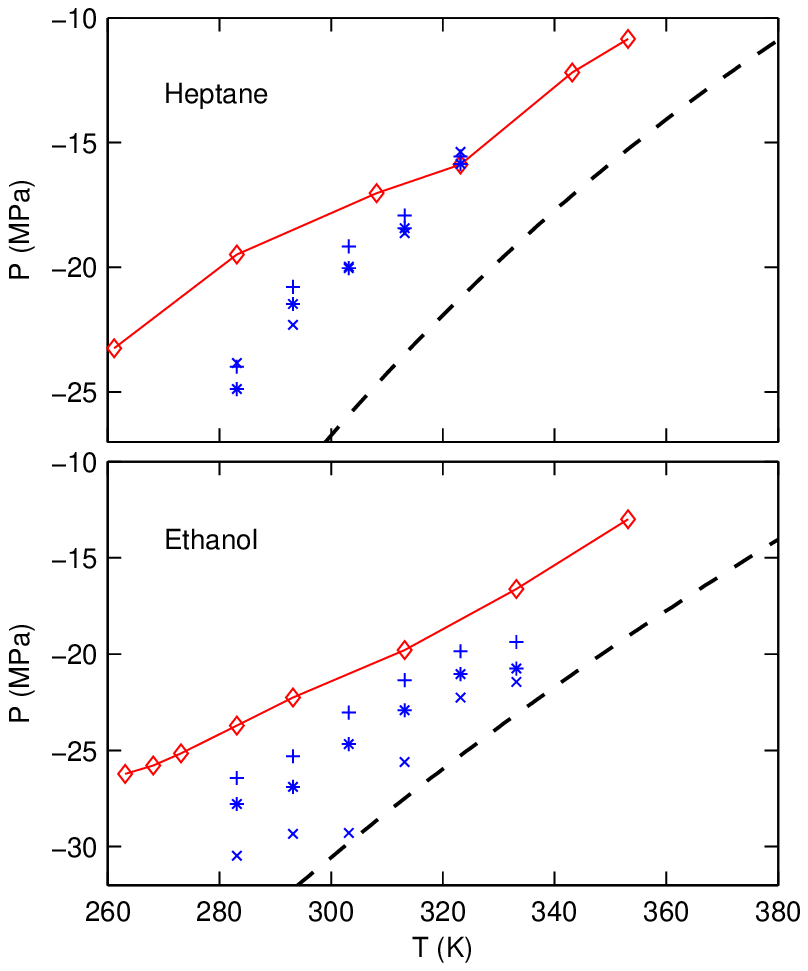}
  \caption{\label{suppl:fig:cavitation_pressures_all_extrapols} Same as in
    Figs.~1 and
    \ref{suppl:fig:nucleation_rates_other} for one of the two FOPH series of
    measurements, but showing the results from the three functions
    $P_1$ (+ markers), $P_2$ ($\ast$ markers) and $P_3$ ($\times$
    markers).}
\end{figure}

\begin{figure}
  \centering
  \includegraphics{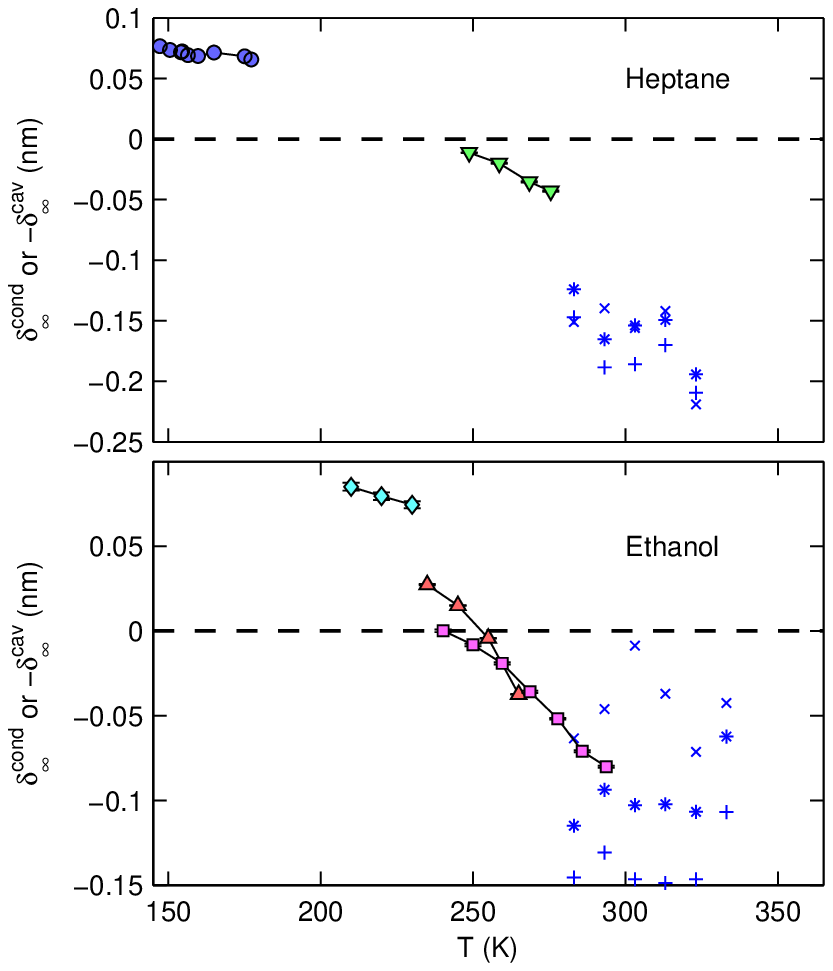}
  \caption{\label{suppl:fig:tolman_order1_all_extrapols} Same as in
    Figs.~2 and
    \ref{suppl:fig:tolman_order1_other} for one of the two FOPH series of
    measurements, but showing the results from the three functions
    $P_1$ (+ markers), $P_2$ ($\ast$ markers) and $P_3$ ($\times$
    markers).}
\end{figure}

\begin{figure}[tttt]
  \centering
  \includegraphics{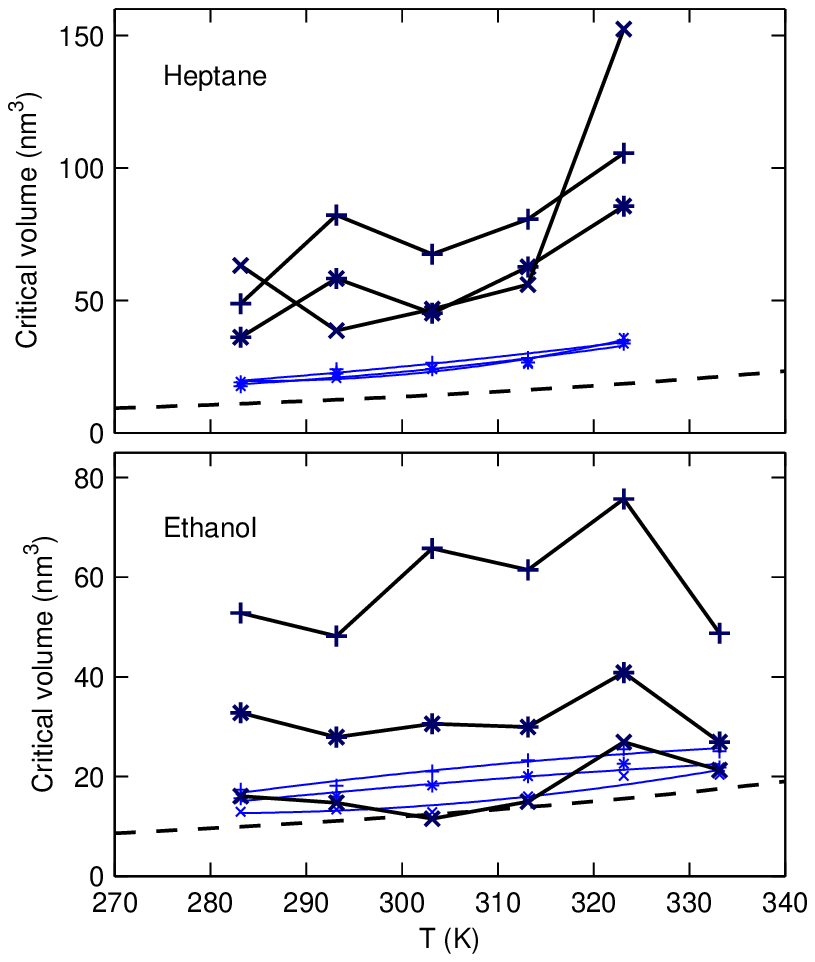}
  \caption{\label{suppl:fig:critical_volumes_all_extrapols} Same as in
    Figs.~3 and
    \ref{suppl:fig:critical_volumes_other}, but showing the results from the
    three functions $P_1$ (+ markers), $P_2$ ($\ast$ markers) and
    $P_3$ ($\times$ markers).}
\end{figure}

\begin{figure}[hhhh]
  \centering
  \includegraphics{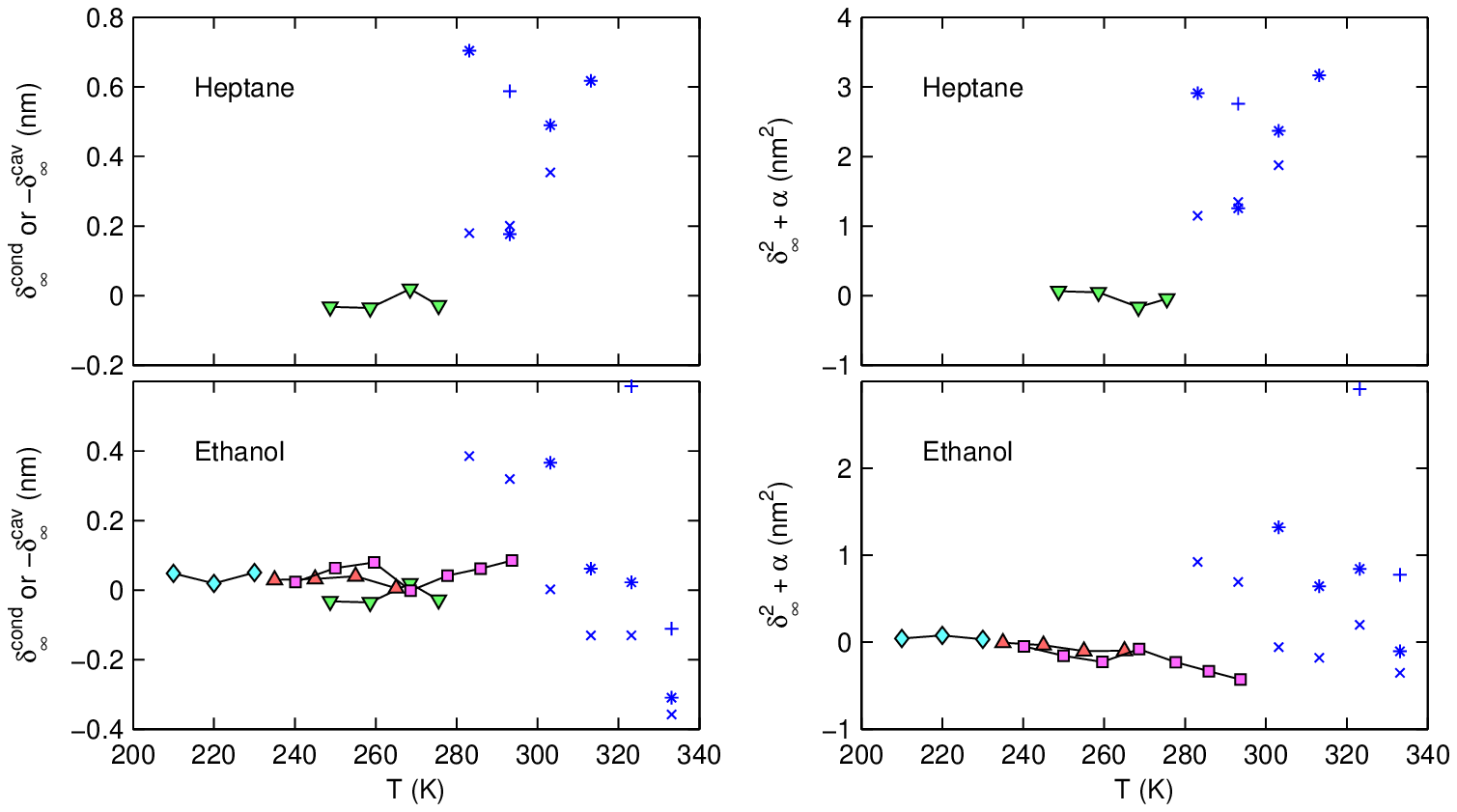}
  \caption{\label{suppl:fig:tolman_order2_all_extrapols} Same as
    Figs.~4 and \ref{suppl:fig:tolman_order2_other},
    but including the cavitation points with polynomial $P_1$ (+
    markers), $P_2$ ($\ast$ markers) and $P_3$ ($\times$ markers).}
\end{figure}

\section{CNT$\boldsymbol{_2}$ for cavitation}
\label{suppl:sec:order2_cav}

Cavitation data have been omitted in the CNT$_2$ analysis.  The reason
is that increasing the order also increases the error bars, and the
errors induced by the $P(U)$ extrapolation (see above) lead to a large
uncertainty in $\delta_\infty$ and $\delta_\infty^2 + \alpha$.
Moreover, the values of the cavitation pressures and of the critical
volumes are such that the solving of the second order polynomial to
obtain the solutions in \myeq{suppl:eq:order2_sols} sometimes gives no real
roots.  \myfig{suppl:fig:tolman_order2_all_extrapols} shows the cavitation
analysis for the points that do have a solution, along with the
condensation experiments.  The systematic error induced by the choice
of the $P(U)$ relation is typically of the order of the range of the
$y$ axis of the graphs for both $\delta_\infty$ and $\delta_\infty^2 +
\alpha$.

\section{Estimate of the statistical error bars}
\label{suppl:sec:error_bars}

Whenever it was possible, we have put statistical error bars on the
quantities $\delta_\infty$ and $V_{\text e}^*$ from CNT$_1$, $V_{\text
  e}^*$ from the NT, and $\delta_\infty$ and $\delta_\infty^2 +
\alpha$ from the CNT$_2$.  This section describes how they were
calculated.

\subsection{Fiber-optic probe hydrophone experiments}

In the fiber-optic probe hydrophone (FOPH) experiments, the error bars
in $\delta_\infty^\text{cav}$ have been estimated on repeated
experiments at a single temperature, $T = 293$~K.  Between two
cavitation pressure measurements, the fiber was removed and
repositioned at the acoustic focal point.  The results were dispersed
with a standard deviation of 0.4~MPa, which we took as the error bar
for all temperatures.
To complement this statistical error, we also display, in the Letter
and here, the FOPH measurements in heptane and ethanol for two series
of temperatures (for each liquid).  When switching to a new series,
the fiber was cleaved to renew its end facet, which leads to an
additional uncertainty in the measurements.

The critical volumes from the nucleation theorem $V_\text{e}^*$ rely
on a nonlinear fit of the probability to cavitate $\Sigma(U)$, with
one of the parameters, $\xi$, representing the ``width'' of the
$\Sigma(U)$ curve that has an ``S'' shape~\cite{arvengas11a}.  The
error bar on $\xi$ has been estimated for each liquid for a single
temperature.  Assuming that the experiment gave a set of points
$\{(U_\text{exp}, \Sigma_\text{exp})_i\}$, we have generated
numerically several sets $\{(U_\text{exp}, \Sigma_\text{num})_i\}$ and
fitted an S-curve on each of them, thus providing us with a standard
error on $\xi$.  The generation of a given point $(U_\text{exp},
\Sigma_\text{num})$ is done by taking for $\Sigma_\text{num}$ a random
value corresponding to the average of $N$ values $\sigma_j \in \{0,
1\}$ with a probability of obtaining 1 equals to $\Sigma_\text{exp}$.

\subsection{Condensation experiments}

The condensation data typically correspond to measurements of the
nucleation rate $J$ for different values of the supersaturation $S$,
and of the temperature $T$.  For a given temperature, the $S(J)$ curve
is expected to be a portion of a line, and such a fit is indeed
performed to obtain the critical volume $V_{\text e}^*$.  We extract
two statistical errors from that fit:
\begin{itemize}
\item The standard error on the slope.
\item A standard error on the average nucleation rate for the given
  temperature.  When plotting the $S(J)$ fit function together with
  the experimental points, these are dispersed around the fit
  function.  Assuming that there is an error in the $J$ measurements,
  but that $S$ is known precisely, this allows to estimate an overall
  statistical error in a single $J$ measurement, $\Delta J_0$, from
  the deviation of the experimental $J$ to the fit function.
\end{itemize}
These made possible the calculation of standard errors on the
following quantities:
\begin{itemize}
\item The first error above is used to get the statistical error on
  $V_{\text e}^*$ (from nucleation theorem).
\item Since we used, for a given temperature, the average value (over
  all $S$) of $J$ as an input in the $\delta_\infty$ and
  $V_\text{e}^*$ formulas, the standard errors on these quantities
  simply derive from the standard error $\Delta J = \Delta J_0 / \sqrt
  N$, with $N$ the number of points for the fit.
\item The $\delta_\infty$ and $\delta_\infty^2 + \alpha$ quantities
  both depend on the two parameters $J$ and $V_\text{e}^*$.  We have
  noted that in most of the cases the error due to $J$ in
  $\delta_\infty$ and $\delta_\infty^2 + \alpha$ is at least 10 times
  smaller than the error due to $V_\text{e}^*$.  We have therefore
  only included the standard error due to the critical volumes in
  $\delta_\infty$ and $\delta_\infty^2 + \alpha$.
\end{itemize}

\section{Second order approximation for the surface tension
  $\boldsymbol{\sigma_{\text s}}$}

\subsection{Validity}
\label{suppl:sec:gtkb_approx}

Eq.~(9) is derived by Taylor expanding and integrating
the Gibbs-Tolman-Koenig-Buff (GTKB) equation (Eq.~(26) in
\myref{troster12}) to second order in $1/R_\text{s}^*$, by neglecting
terms such as $\delta_\infty^3/(R_{\text s}^*)^3$ and
$\alpha^2/(R_{\text s}^*)^4$.  For each pair of values
$(\delta_\infty, \delta_\infty^2 + \alpha)$ from the experiments, we
have compared the value of $\sigma_\text{s}$ from
Eq.~(9) with the full numerical integration of the GTKB
equation. For the 34 points used to plot Figs.~4 and \ref{suppl:fig:tolman_order2_other}, the maximum relative error is 3.7~\%, and 29 points give an error of less than 1~\%, which makes
Eq.~(9) an excellent approximation for realistic ranges
of parameters for condensation.

\begin{figure}[tttt]
  \centering
  \includegraphics{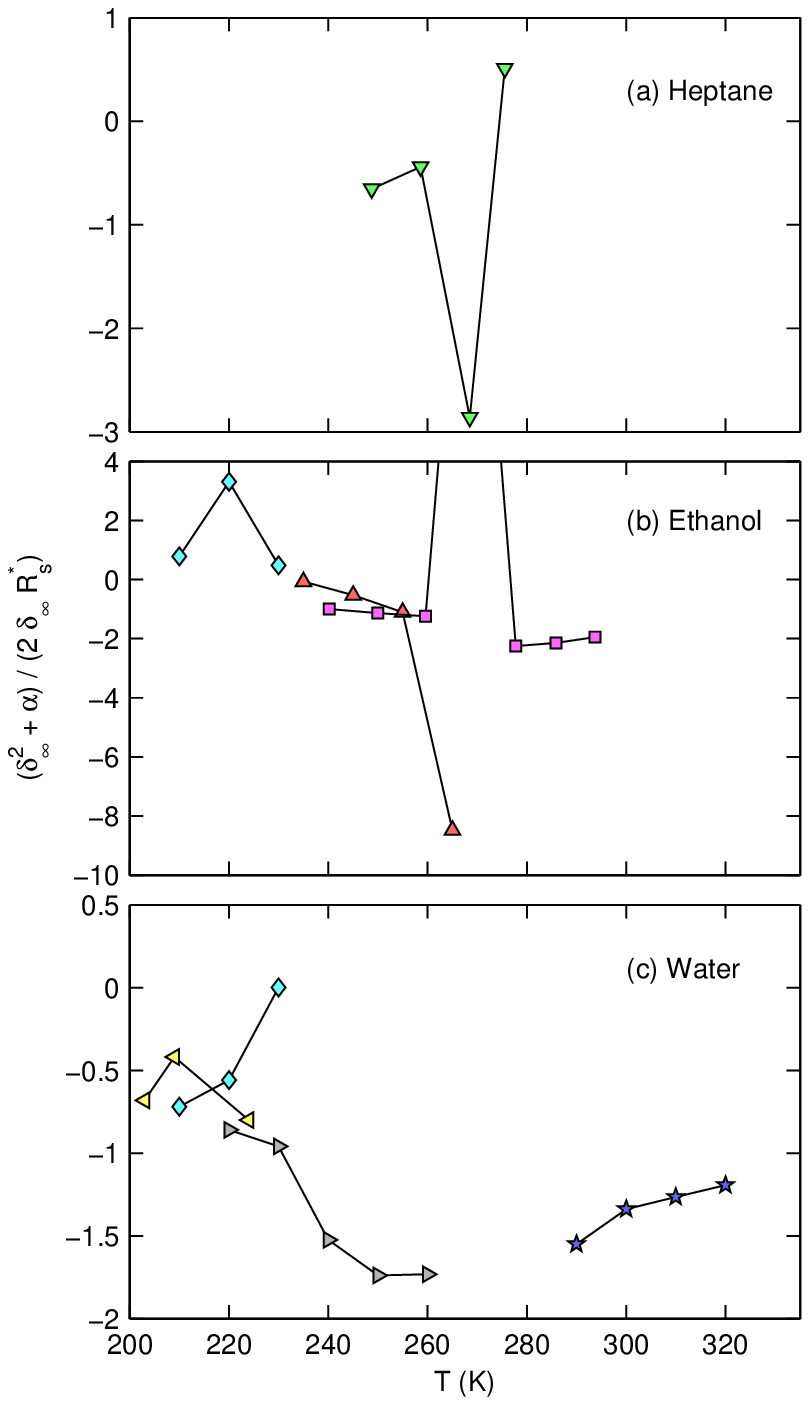}
  \caption{\label{suppl:fig:orders_ratio} Ratio between the second order
    term $(\delta_\infty^2 + \alpha) / (R_\text{s}^*)^2$ and the first
    order term $2 \delta_\infty / R_\text{s}^*$.  The point out of the
    graph for ethanol is at a height $(\delta_\infty^2 + \alpha) / (2
    \delta_\infty R_s^*) = 269$.}
\end{figure}

\subsection{Comparison of the two terms in the
  $\boldsymbol{\sigma_{\text s}}$ expansion of the
  CNT$\boldsymbol{_2}$}
\label{suppl:sec:orders_ampl}

With the beginning of the use of models including a second order
related to the rigidity of the interface, some debate emerged about
the relative amplitudes of the two terms $2 \delta_\infty /
R_\text{s}^*$ and $(\delta_\infty^2 + \alpha) / (R_\text{s}^*)^2$.  In
simulations, models that include the second term only, or that include
the two terms~\cite{block10,baidakov14} have been tested.  For the
experiments, their ratio, extracted from the condensation
measurements, is shown in \myfig{suppl:fig:orders_ratio}.  Overall, we found
that the two quantities are often of the same order of magnitude.
Sometimes, the ratio takes values much larger than 1, but this only
happens when $\delta_\infty$ is very close to 0, so that these points
may have large error bars, because of the error in the $\delta_\infty$
values.  Therefore, we expect that a model of a surface tension
varying as $\sigma_\infty/\sigma_\text{s} = 1 + C/(R_\text{s}^*)^2$,
with $C$ a constant, would display inconsistencies similar to those
found with the CNT$_1$ model.

\section{Fitted CNT$\boldsymbol{_2}$ parameters for ethanol}

Considering the condensation data only, we obtained an overall
agreement of all the data in the sense that the points are closer to
master curves for $\delta_\infty(T)$ and $(\delta_\infty^2 +
\alpha)(T)$ in the CNT$_2$ than what was found for $\delta_\infty$ in
the CNT$_1$.  The best results are obtained for ethanol.  We have
fitted in this case these functions by simple expressions that can be
used as empirical functions, namely a constant $\delta_\infty$ and a
linear $\delta_\infty^2 + \alpha$:
\begin{equation}
  \label{suppl:eq:order2_param_fit}
  \delta_\infty^2 + \alpha = A (T - T_{\text{ref}}) + B
\end{equation}
With $T_{\text{ref}} = 298.15$~K, the fit parameters are
$\delta_\infty = 0.04095$~nm, $A = -0.005460$~nm$^2$/K and $B =
-0.3352$~nm$^2$.

% \bibliography{cavitation_condensation_suppl}
% \bibliographystyle{ieeetr}